# The role of 'living laboratories' in unlocking the potential of low-carbon energy technologies on the UK's journey to net-zero

Zhong Fan[a], Jun Cao[b], Taskin Jamal[c], Chris Fogwill[d], Cephas Samende[a], Zoe Robinson[a], Fiona Polack[e], Mark Ormerod[a], Sharon George[a], Adam Peacock[a], David Healey[a]

[a] *Keele University*
[b] *Luxembourg Institute of Science and Technology (LIST)*
[c] *Department of Electrical and Electronic Engineering, Ahsanullah University of Science and Technology, Dhaka, Bangladesh*
[d] *Cranfield University*
[e] *University of Hull*

**Abstract:** Together, renewable energy systems (RES) and distributed and digitized 'smart' energy networks (SEN) provide opportunities to maximize energy efficiency, reduce transmission losses and drive down greenhouse gas emissions. Yet, such integrated Smart Local Energy Systems (SLES) are in the early stages of development and the technologies that underpin them lack testbeds where they can be developed and tested in a real-world environment. Here we demonstrate the potential role of one of Europe's largest 'at scale' multi-vector Smart Energy Network Demonstrator - SEND, developed within a 'living laboratory' setting that provides a 'blueprint' for the development and testing of low-carbon energy technologies on the UK's journey to net zero.

## 1. Introduction

Climate change continues to be a daunting challenge and is the single biggest global threat to ever face us. Globally, the energy sector (electricity, heat and transport) is responsible for about 73% of greenhouse gas emissions driving climate change [1]. The UK government has set a commitment to achieve net-zero carbon emissions by 2050 with decarbonisation of the energy sector as priority areas [2]. Thus, urgent development and deployment of low-carbon energy technologies are needed in the energy sector to: (i) better understand how we generate, distribute and consume energy and (ii) reduce our reliance on fossil fuels to significantly reduce the greenhouse gas emissions.

Although the deployment of low-carbon energy technologies is key to mitigating climate change, uncertainty about a technology's performance and the integration of different technologies is an important barrier to adoption [3]. Living labs can overcome such barriers by providing a real-world environment where such new low-carbon energy technologies and interventions can be researched, developed and trialled prior to larger-scale deployment. Without evidence that the low-carbon energy technologies being deployed have been tested and evaluated using a real-world environment, it is unlikely that the public and other stakeholders will risk adopting and supporting them [4]. Further, the power in the living lab approach in a higher education setting, is in the potential to maximise the impact across both the education and research missions of a university, as well as in the benefits of working with businesses and wider community with multiple stakeholders represented in discussions from the outset.

In this comment article, we demonstrate the potential role of Keele University's Smart Energy Network Demonstrator - SEND, developed within a 'living laboratory' setting to provide the 'blueprint' for research and testing of low-carbon energy technologies and the development of a Smart Local Energy System as part of the UK's journey to net zero.

## 2. Overview of the SEND

Funded by the European Regional Development Fund and Department for Business, Energy and Industrial Strategy (BEIS), the Keele University SEND was established to create Europe's largest 'at scale' multi-vector smart energy demonstrator on the Keele campus. The demonstrator provides a platform that allows energy generation, distribution, storage, forecasting and energy balancing to be intelligently carried out across different energy sources using the Keele University campus as a genuine 'living laboratory' as shown in Figure 1.

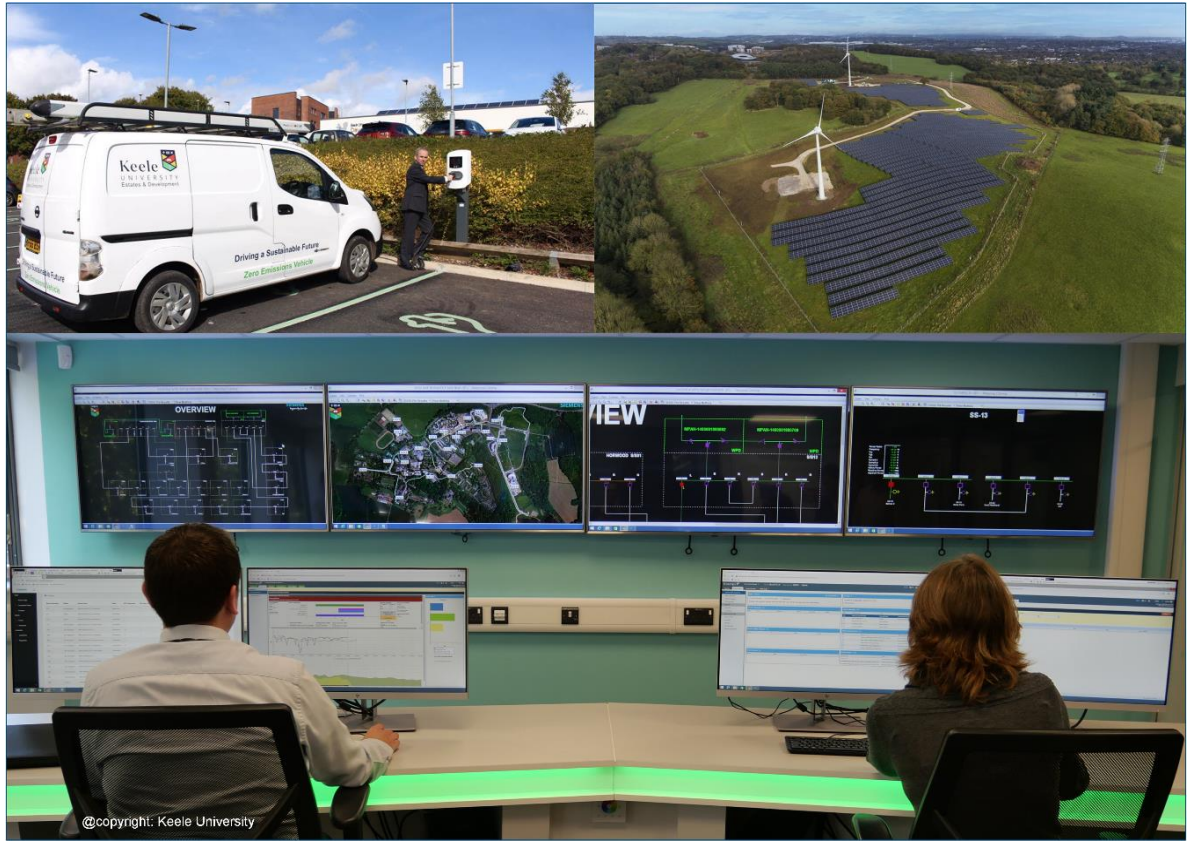

Figure 1: Overview of the SEND Living Laboratory at Keele University - Electric vehicle charging points (top-left), renewable energy generation using wind and solar (top-right), and SEND control room (bottom) (copyright: Keele University)

Keele University owns and operates all of the utilities on its 600-acre campus, thus creating a single owner/operator environment across power, gas, heat, telecoms, water and drainage. The campus operates at the scale of a small town with 5,000 residents and more than 12,000 staff and students on campus per day and includes over 100 domestic properties for staff housing, as well as over 3000 student residential bedrooms, catering and leisure facilities, and office, teaching and laboratory environments. Annual energy demand across the campus is circa 63 GWh.

Being a private network, SEND is able to integrate existing gas, water, and electricity networks to form a digitally managed, flexible and extensible smart system. Keele University has worked with Siemens and Equans to install the following infrastructure on the SEND network: 1500 smart meters (electric, heat and gas), 5.5MW Solar farm, 1.9 MW wind turbines, 1 MW battery storage, upgrade of 25 substations, Spectrum distribution management system, decentralized energy management system (DEMS), Mindsphere (IoT cloud platform), SEND digital twin, IoT smart appliance control, over 20 EV charging points, and many residential, commercial and industrial controllers.

The SEND system is an ideal testbed to explore how intelligent generation and distribution of electrical energy can be managed in a SLES environment. SEND enables the campus system operators to control how the generation from the renewable energy sources is self-consumed in the Keele distribution network, feed-in of the excess generation to the grid and optimize the energy use on campus by measuring the grid carbon intensity. This is an innovative scheme in the way that the researchers and system operators can test novel leading-edge algorithms, methodologies and resolutions for energy distribution in a closed environment with energy-conscious consumers. The Living Lab approach also provides the opportunity for essential research to better understand the real use patterns and perceptions of smart energy system components of diverse energy users, and exploration of challenges at the socio-technical interface, as well as offering the opportunity to increase the energy transition literacy of the campus community.

## 3. The SEND System Architecture

Figure 2 shows the whole SEND system architecture, including Infrastructure, Control and Management, and Dataset. Central to the SEND is the control centre and digital twin, which facilitates distribution system management, power system simulation, distributed energy optimization and augmented reality as described below:

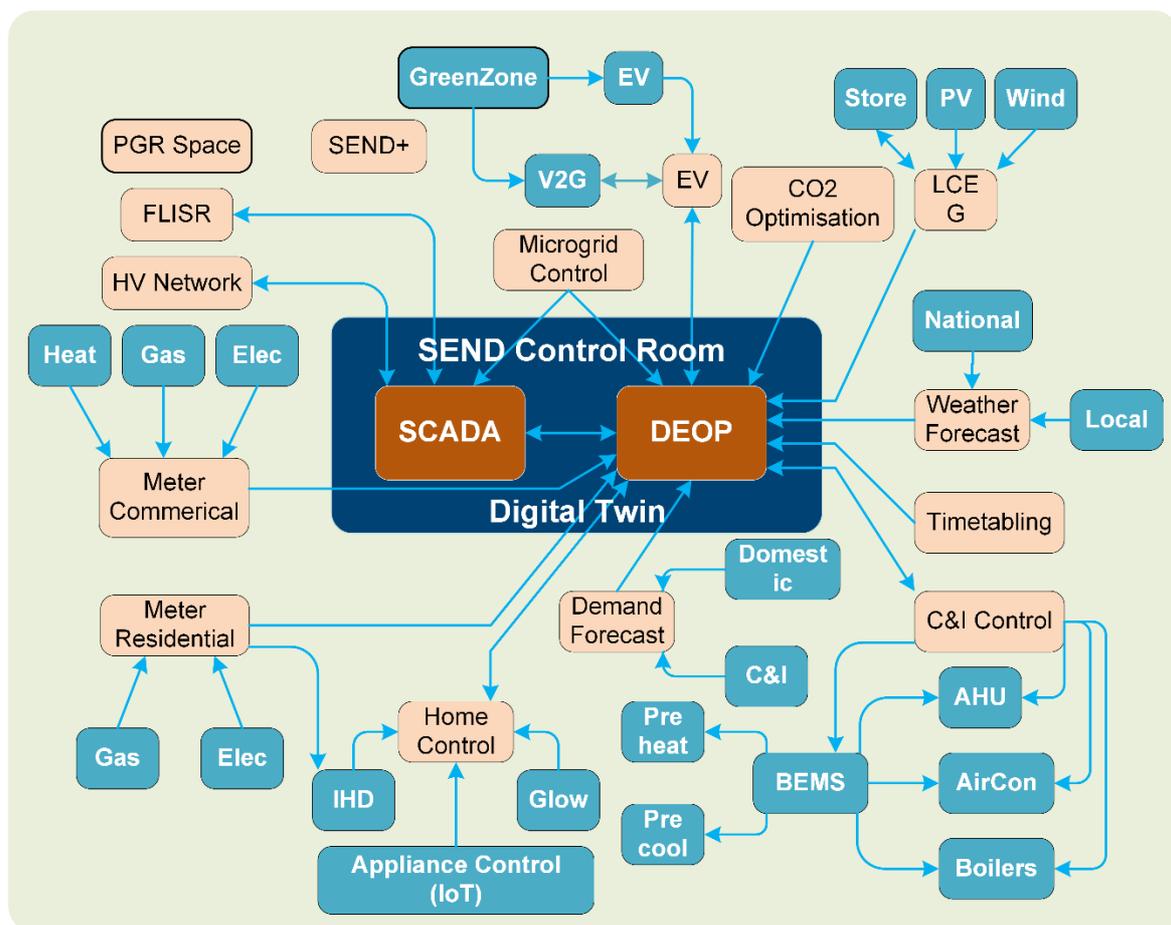

Figure 2: SEND system architecture

*Advanced Distribution Management System*: The advanced distribution management system is facilitated by the Siemens Spectrum software. It controls Keele's medium voltage (11 kV) network, conveying real-time substation data, predicting possible demand constraints and is capable of intelligent healing the network in the event of outages. It enables the SEND network to establish the supervisory control and data acquisition (SCADA), thus assisting system controllers. Further, it allows reconfiguring, tweaking, and testing the impact of planned and possible network modifications before they are committed.

*Power System Model and Simulation*: The SEND system utilizes the software tool PSS Sincal to model, simulate, plan and analyze Keele's power network using live or offline data. Technical information coupled with geographic information from the Spectrum and SCADA can be fed to PSS Sincal to examine the variabilities in the power networks according to various scenarios and explore the technical issues such as power flow analysis, optimal routing, short circuit analysis, etc. A broad range of multi-vector energy systems comprising of renewable energy resources, storage, the conventional resources such as gas turbines, combined heat and power, hot water and loads are studied by the PSS simulator.

*Distributed Energy Optimizer Platform*: The whole dataset of SEND is managed by a Distributed Energy Optimizer Platform (DEOP), a cloud-based open and flexible platform that uses sophisticated algorithms to control various campus assets. DEOP can include third-party algorithms developed in a compatible programming language, e.g., Python. DEOP can manage controllable assets according to an optimization logic configured by the user. The first goal for optimization in SEND is to minimize the $CO_2$ in the grid, which can be pursued by taking into account all the emission factors of generation assets and consumption assets.

*Data Visualization and Analysis*: DEOP can activate the Demand Side Management (DSM) facility of the load flexible campus buildings, which can either increase the building load, balance the on-site renewable generation or reduce the load in response to one of the two options, day-ahead or ad hoc. Figure 3 showcases an example of DSM mechanism for $CO_2$ minimization in a flexible building at Keele on September 29, 2021. The nominal AC power capacity from the PV systems is 4.4 MW. Zooming into high-resolution data, weekly and daily data can be read from Figure 4 (Sept. 21-27, 2021). During this week, the total energy demand was 202.11 MWh. The amount supplied by the PV system was 86.260 MWh, from where the consumers directly consumed 65.893 MWh, and the rest amount of 20.374 MWh was fed into the grid. These account for 32.599% of self-sufficiency (SS) and 76.388% of self-consumption (SC) from PV systems.

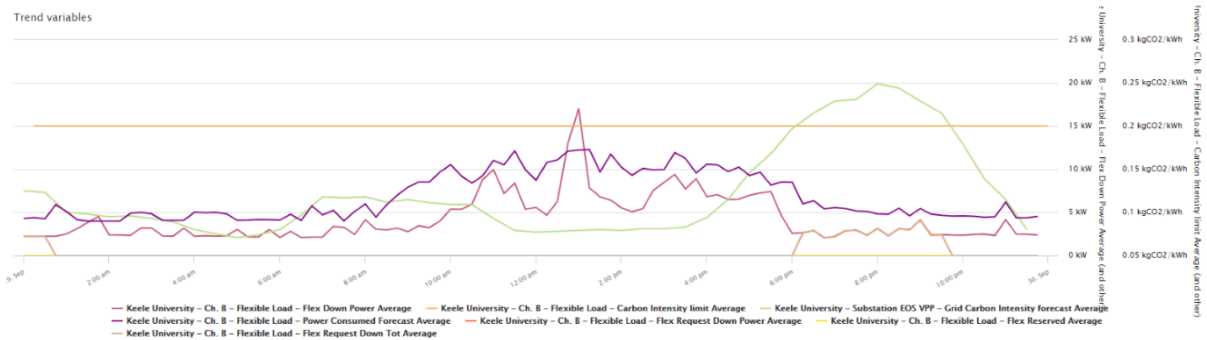

Figure 3: Example of DSM mechanism occurring for $CO_2$ minimization in a flexible building

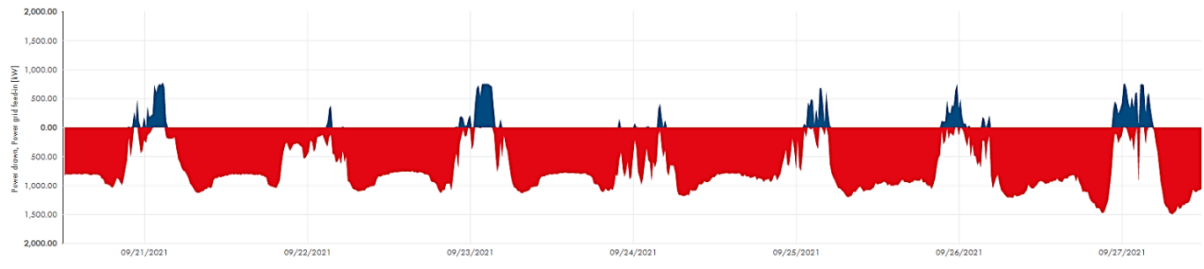

(a) 09/21/2021 - 09/27/2021

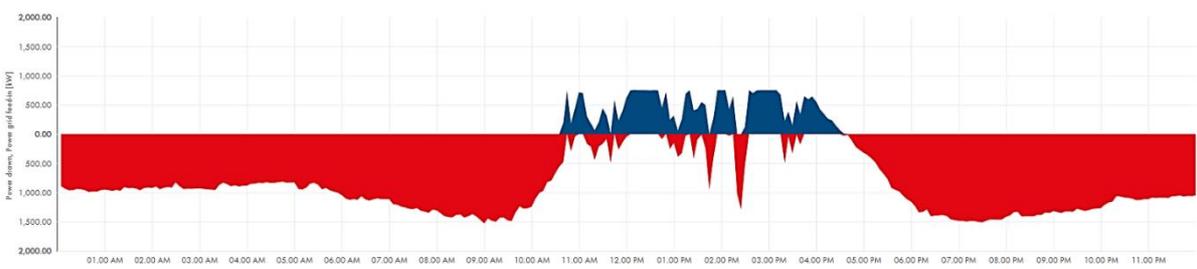

(b) on 09/27/2021

Figure 4: Weekly and daily distribution of grid feed-in and consumption (red colour – power drawn from the grid, blue colour – feed-into the grid from excess generation in the campus)

*Augmented Reality*: The augmented reality disseminates system data in all possible ways a user wants, including the heatmap to show system efficiency, in-depth system statistics, network, renewable energy feed using a graphical dashboard, etc. as shown in Figure 5.

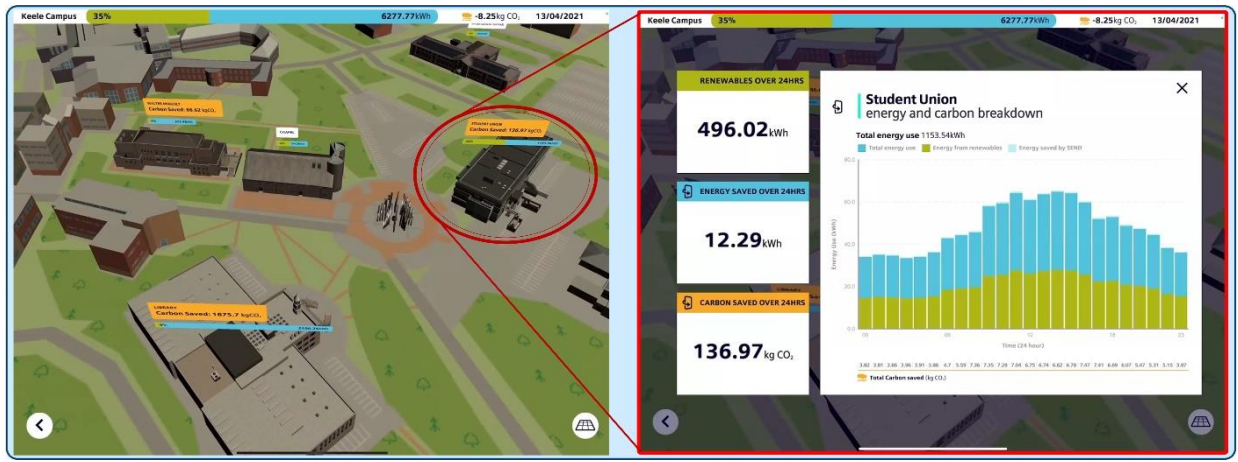

Figure 5: Augmented reality showing the amount of renewables used, energy and carbon saved over 24 hrs at different facilities across Keele campus on 13/04/2021. For example, 496.02 kWh of renewables was used, 12.29 kWh of energy was saved and 136.97kg $CO_2$ of carbon was saved at the Student Union building.

## 4. Potential Roles of the Keele University's Living Laboratory

The decarbonization, decentralization, and digitization of the energy sector requires balancing environmental, economic, and social concerns. This is challenging due to the risk in passive trade-offs between equally critical priorities. These trade-offs can be explored through the living laboratory to develop a better understanding of the implications of decision making around these trade-offs and ways to optimise the system operation. With the campus buildings and infrastructure ranging from late Victorian to ultra-modern, the opportunity to explore energy management approaches at the campus scale and examine the system efficiency and the resulting carbon emissions in detail inside this controlled environment provides a remarkable opportunity.

The digital twin [5] within SEND is based on models of the physical components and takes input from demands that are fed into the system, other exogenous data (such as weather forecast or actual weather data, various IoT sensor data) to investigate different "what-if" scenarios. We are also investigating the combination of physics-based energy network models and advanced explainable AI techniques (e.g., deep neural networks and reinforcement learning) [6] to facilitate the data-driven optimal design and real-time operation of future net-zero energy systems, by harnessing the power of live data from the SEND platform.

The living laboratory's capabilities allow the engineering community to design a roadmap for Grid 2.0 in the UK, where grid digitalization and decentralization demonstrate the possibilities for energy supply chain decarbonization and ensure supply security. The UK government (as well as Keele University itself) has set aggressive ambitions to reach net-zero emissions, and in this regard, the Grid 2.0 supported SLES can assist in meeting these objectives. Keele's decentralized energy generation, distribution and management system based on advanced modern technologies at the local level, creates learning on the critical path to net-zero while also providing significant commercial potential and opportunities for the development of cutting-edge energy technologies. The innovative testbed encourages and welcomes research and implementation opportunities based on machine learning and big data analytics. For example, working with industry partners, we have been researching and trialling the novel concept of peer-to-peer (P2P) energy trading on this testbed [7], without the requirement for regulatory change as for a public network. Electricity trading on a P2P basis has the potential to empower prosumers and consumers, resulting in increased renewable energy development and system flexibility. Demonstrating effectiveness and learning from deployment on a private living lab network can help unlock the barriers to wider-scale implementation.

The living laboratory allows for innovation and infrastructure development for regional and national testing and pilot programs for government projects. It provides space and opportunities for large to small, multinational and multisectoral companies to test and demonstrate how sustainable cities and communities should look. Towns, organizations and companies can help decarbonize their operations and supply chains by learning from the pilot projects and real-life demonstrations using the living lab. By extending the hand of partnership and cooperation, the living laboratory at Keele can strengthen the means of implementation and revitalize the global partnership for sustainable development.

## 5. Conclusion

Learnings from the living laboratory allow improvement and enablement of initiatives all around the world, not only in the UK. The technologies developed can also be scaled up to other big demonstrators and SLES systems. This will pave the way for knowledge sharing on how communities can benefit from the low-carbon energy future. Ongoing research activities on the living lab include using SEND data on applying advanced artificial intelligence and blockchain technology to SLES, study of consumer behaviour and perception of novel SLES technologies, user-centred design of SLES, as well as new materials for smart energy. Leveraging the SEND platform for scaling up and collaboration, we have been working with many institutions in UKRI funded projects such as EnergyREV [8], Hydex [9], and Zero

Carbon Rugeley [10], to play a major role in researching and developing low-carbon energy technologies as part of the UK's journey to net-zero.

**Acknowledgements:**

This work is partly supported by the SEND project (grant ref. 32R16P00706) funded by ERDF and BEIS as well as EPSRC EnergyREV (EP/S031863/1). We also thank Ash Dean, Matt Dean, Ian Shaw, Phil Butters, Mark Turner, Julian Read, and Ash Hulme for their support.